\documentclass{article}

\usepackage{arxiv}

\usepackage[utf8]{inputenc} 
\usepackage[T1]{fontenc}    
\usepackage{hyperref}       
\usepackage{url}            
\usepackage{booktabs}       
\usepackage{amsfonts}       
\usepackage{nicefrac}       
\usepackage{microtype}      
\usepackage{hyperref}
\usepackage{cleveref}
\usepackage{graphicx}
\usepackage[square, sort, comma, numbers]{natbib}

\usepackage{caption}
\usepackage{wrapfig}
\usepackage{listings}

\usepackage{fontawesome5}

\title{AEStream: Accelerated event-based processing with coroutines}

\author{ \href{https://orcid.org/0000-0001-6012-7415}{\includegraphics[scale=0.06]{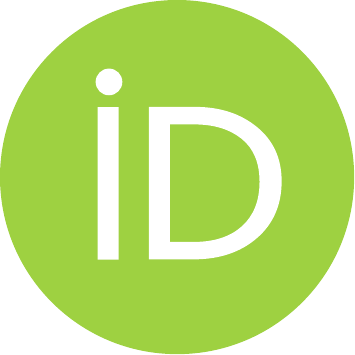}\hspace{1mm}Jens E.~Pedersen} \\
	Department of Computer Science\\
	KTH Royal Institute of Technology\\
	Stockholm, Sweden \\
	\texttt{jeped@kth.se} \\
	\And
	\href{https://orcid.org/0000-0001-5998-9640}{\includegraphics[scale=0.06]{orcid.pdf}\hspace{1mm}Jörg Conradt} \\
	Department of Computer Science\\
	KTH Royal Institute of Technology\\
	Stockholm, Sweden \\
	\texttt{conr@kth.se} \\
}

\renewcommand{\undertitle}{Under review}
\renewcommand{\shorttitle}{AEStream}

\hypersetup{
pdftitle={AEStream: Accelerated event-based processing with coroutines},
pdfsubject={q-bio.NC, q-bio.QM},
pdfauthor={David S.~Hippocampus, Elias D.~Striatum},
pdfkeywords={First keyword, Second keyword, More},
}

\begin{document}
\maketitle

\begin{abstract}
Neuromorphic sensors imitate the sparse and event-based communication seen in biological sensory organs and brains.
Today's sensors can emit many millions of asynchronous events per second, which is challenging to process on conventional computers.
To avoid bottleneck effects, there is a need to apply and improve concurrent and parallel processing of events.

We present AEStream: a library to efficiently stream asynchronous events from inputs to outputs on conventional computers.
AEStream leverages cooperative multitasking primitives known as coroutines to concurrently process individual events, which dramatically simplifies the integration with event-based peripherals, such as event-based cameras and (neuromorphic) asynchronous hardware.
We explore the effects of coroutines in concurrent settings by benchmarking them against conventional threading mechanisms, and find that AEStream provides at least twice the throughput.
We then apply AEStream in a real-time edge detection task on a GPU and demonstrate 1.3 times faster processing with 5 times fewer memory operations.
\end{abstract}


\section{Introduction}

Current event-based megapixel resolution cameras can emit tens of millions of events every second~\citep{Tayarani-Najaran_Schmuker_2021}.
Processing the asynchronous events is ideally done in parallel compute substrates, such as the neuromorphic platforms SpiNNaker~\cite{Furber_Galluppi_Temple_Plana_2014, Mayr2019SpiNNaker2A} and BrainScaleS~\cite{Pehle2022TheBA}.
Neuromorphic hardware is already operating several orders of magnitude faster and with far less energy compared to conventional von Neumann architectures.
And promising new perspectives in physical and material sciences indicate that neuromorphic technologies are only in the very early stages~\cite{Zhu_Zhang_Yang_Huang_2020}.

Despite these promises and perspectives, digital computers still prevail and are also experiencing impressive growth.
The number of operations per second performed in Graphical Processing Units (GPUs), Tensor Processing Units (TPUs), and Application-Specific Integrated Circuits (ASICs) are developing at an almost exponential rate, mostly because they compute in \textit{parallel}~\cite{Dally_Keckler_Kirk_2021}.
Unfortunately, digital computers use centralized, synchronous architectures that align poorly with asynchronous data, such as events.
To operate multiple compute cores in parallel, they require time to synchronize shared memory, which incurs significant overhead.
Amdahl concretized this overhead by stating that the theoretical speedup of parallelizing work is limited by the amount of work that can actually be parallelized \cite{Amdahl_1967}.
This seemingly trivial insight is particularly sobering as we approach the physical limits of digital systems, but also on a more practical level when operating locks and peripheral devices---such as event-based sensors.

The silver lining in this conundrum arrives from the fact that Amdahl's law is bounded only by shared bottlenecks.
Theoretically, no such bottlenecks are required in event-based processing because events are independent.
Modelling computation as independent functions has deep roots in the origins of computer science, namely lambda calculus and, more broadly, functional programming.
\textit{Functional} descriptions operate on individual atoms (such as events), which is a useful abstraction regardless of whether we are working with neuromorphic or conventional hardware.
One way to operate with functional atomic operations is in the form of \textit{streams}, where data packets ``flow'' through concurrent processing pipelines~\cite{Hughes_2000}.
Von Neumann computers can exploit this ``flow'' to assign cores in cooperation with each other, known as cooperative multitasking, and thereby increase the effects of parallelization.
The new C++20 standard recently introduced coroutines to support exactly this kind of concurrency~\cite{CoroutinesCppReference}.
Coroutines are, simply put, functions that do not require synchronization to pass on data.
In the context of event-based processing, they allow us to operate directly on the level of events, in contrast to buffers or other synchronization mechanisms that would normally be required to exploit parallelization benefits.
This further simplifies integration with neuromorphic sensors and hardware, since they both communicate with events.

In this work, we demonstrate a method to process event-based data that operates well on conventional hardware, while retaining compatibility with neuromorphic and parallel peripherals.
We begin by revisiting a functional approach to event-based processing and demonstrate how coroutines can double the throughput across synchronization barriers.
We then apply our method to identify edges in a live stream of events using a spiking neural network running on a GPU.
Coupled with the improved throughput mentioned above, we can exploit the parallelism of the GPU to process 1.3 times more data with 5 times fewer memory operations.

Due to the simplicity of the interface, our method has been straight-forward to integrate event-based inputs and outputs, and we demonstrate support for cameras, network protocols, and event-based file representations, as discussed in Section \ref{sec:aestream} and visualized in Figure \ref{fig:architecture}.
Importantly, this includes peripherals such as GPUs, and the neuromorphic hardware platform SpiNNaker.

We implemented our method in the software library AEStream, which we release openly along with all the necessary steps to reproduce our findings at \href{https://jegp.github.io/aestream-paper}{https://jegp.github.io/aestream-paper}.

\section{Address-event representations and processing}
\begin{figure}
\centering
\includegraphics[width=0.5\textwidth]{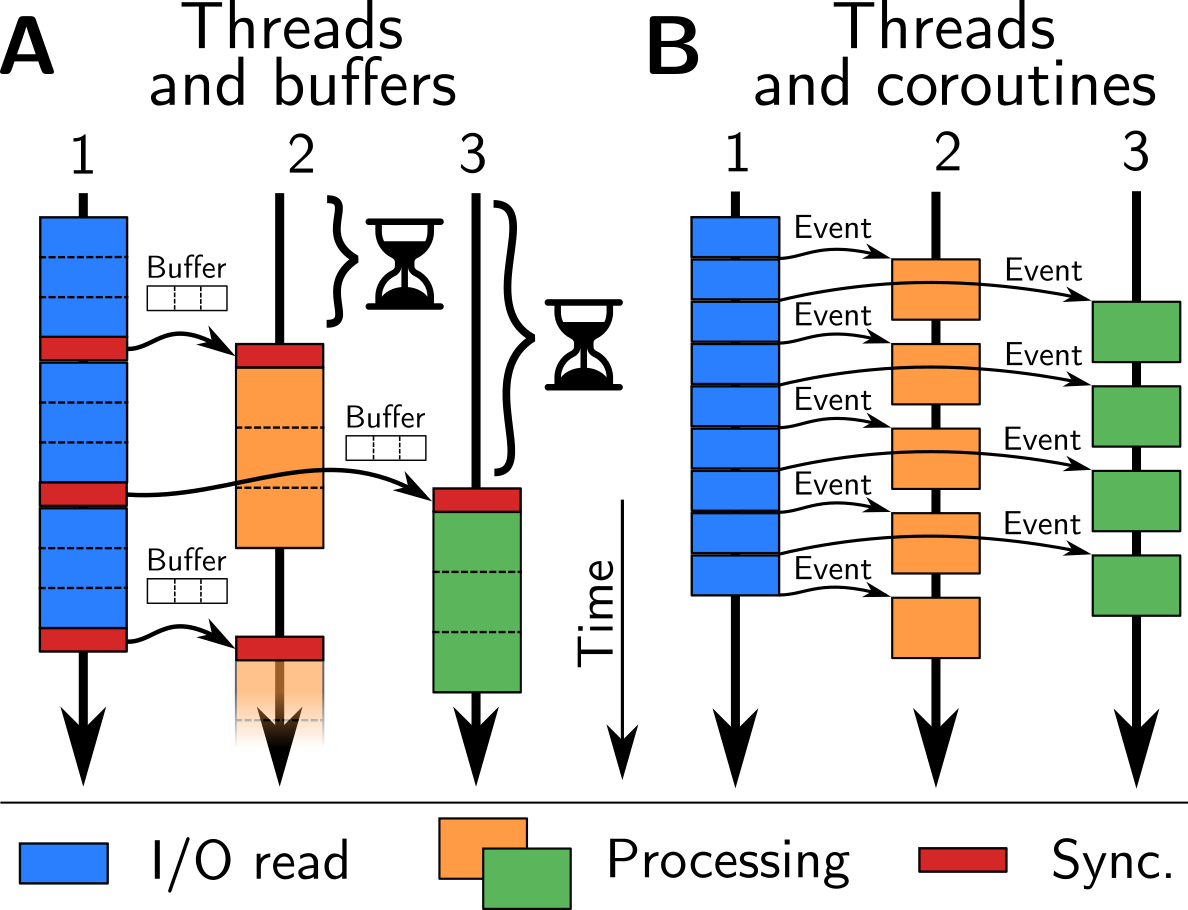}
\caption{
\textbf{(A)} Lock-based synchronization mechanisms fill up buffers from an I/O thread (blue) and release them when they are full to ``activate'' other threads.
\textbf{(B)} Coroutines can \texttt{suspend} execution and transfer control to other functions on the level of singular events with an overhead comparable to a regular function call.
}
\label{fig:coroutine}
\end{figure}
Address-event representations (AER) were first proposed by Carver Mead in 1991 as a low-power, high-throughput method to transfer discrete neuron events, or \textit{spikes}~\cite{Mead1989AnalogVA, Liu2015EventBasedNS}.
Replacing neurons with pixels, AER encoding has been widely used for silicon retinas that exactly emit discrete events, depending on when the light intensity for a singular pixel in the retina changes by a given threshold~\cite{Fukushima1970AnEM, Liu2015EventBasedNS}.
AER formats individual events as singular atoms of spatial, temporal, and polarity information.
Events are typically represented as 4-tuples $(x,y,p,t)$, where $\{x,y\} \in \mathbb{N}$ are spatial coordinates, $t \in \mathbb{N}$ denotes a timestamp, and $p \in \{\texttt{True}, \texttt{False}\}$ indicates positive or negative event polarity, depending on the direction of the luminosity change~\cite{Lichtsteiner_Posch_Delbruck_2008}.

\subsection{Threading and synchronization barriers} \label{sec:sync}
On conventional computers, processing AER data can be viewed as the simple problem of ferrying inputs from a \textit{source} to a \textit{sink} (see Figure \ref{fig:architecture}).
However, effectively communicating with input and output (I/O) devices in modern operating systems requires the use of threads to maximize peripheral throughput~\cite{Silberschatz1983OperatingSC}.
In turn, that requires synchronizing the input and output threads via some form of shared memory, which is typically solved by locking said memory during reading or writing~\cite{Silberschatz1983OperatingSC}.
Locking and unlocking costs time and creates bottlenecks if the input is not providing data fast enough, or if the output is not taking data from the shared memory fast enough.
Figure \ref{fig:coroutine} \textbf{(A)} illustrates the former: thread 2 waits for the IO thread (1) to provide a buffer of events.

Finding efficient ways to operate locks and synchronization barriers is a highly active area of research because they are so ubiquitous.
Approaches to eliminate locks have also been suggested, such as lock-free transactional memory~\cite{Herlihy1993TransactionalMA} and $\mathcal{O}(1)$-time memory sharing with immutable data structures~\cite{Okasaki1998PurelyFD}, but they are ``rarely suitable for practical use'' due to complexity, risk of deadlocks, and excessive use of memory or compute resources~\cite[p. 1]{Fraser2007ConcurrentPW}.

\subsection{Coroutines}
Coroutines were defined by Melvin Conway in 1958~\cite{Knuth1968TheAO} as a way to \textit{pass control} between functions, without the need for centralized synchronization.
Throughout the lifetime of coroutines, there have been several implementations with some ambiguity in terminology, for instance around the use of call stacks.
In the context of this paper, we rely on the C++20 specification and their use of stackless coroutines~\cite{CoroutinesCppReference}.
Coroutines can be stackless because they allow \texttt{suspending} and \texttt{resuming} the execution of a given subroutine at specific points without passing (potentially large) stack frames around.
When a \texttt{suspend} happens, a coroutine stores its execution state and local variables on the heap.
The coroutine can be \texttt{resume}d in the same thread, but, since the state and variables of the coroutine are local, it can even be picked up in any other thread.
If we additionally ensure that the local variables are never changed in the previous, \texttt{resume}d coroutine, we can guarantee that the local memory is exclusive to the new, processing coroutine and, effectively, lock-free.
In theory, this means that coroutines on multicore systems can maximize core utilization without the overhead of locks. 
This is shown in Figure \ref{fig:coroutine} \textbf{(B)} where threads are kept mostly busy without spending time to synchronize, which processes individual events, rather than buffers, faster than the synchronized approach in panel \textbf{(A)}.
Note that coroutines only allow cores to cooperate around and preempt multitasking.
As such, coroutines do not implement parallelism.


\section{Related work}
\newcommand{\faFileO}{\faIcon{file}}
\begin{table}
\begin{tabular}{@{}lllllllll}
\toprule
\textbf{Library}         & \textbf{AEDAT} & \textbf{AEStream} & \textbf{Celex} & \textbf{Expelliarmus} & \textbf{jAER} & 
\textbf{LibCAER} & \textbf{OpenEB} & \textbf{Sepia} \\ \midrule
\textbf{Language}        & Rust           & C++               & C++                & C                     & Java          & 
C/C++            & C++             & C++            \\ \midrule
\textbf{Python bindings} & Yes            & Yes               & No                 & Yes                   & No            & 
No               & Yes             & No             \\ \midrule
\textbf{Inputs}          & \faFileO        & \faIcon[regular]{camera}\hspace{4pt}\faFileO\hspace{4pt}\faWifi  & \faFileO   & \faFileO & \faCamera\hspace{4pt}\faFileO  & \faCamera        & \faCamera\hspace{4pt}\faFileO & \faFileO\hspace{4pt}\faCamera\hspace{2pt}* \\ \midrule
\textbf{Outputs}         & N/A            & \faMicrochip\hspace{4pt}\faFileO\hspace{4pt}\faWifi & \faFileO & \faFileO & \faFileO\hspace{4pt}\faWifi   & N/A      & \faFileO        & N/A            \\ \bottomrule
\end{tabular}
\caption{
An overview of open-source libraries for event-based processing based on the underlying code, python bindings, and native I/O support.
Icons indicate support for GPUs (\faMicrochip), event-based cameras (\faCamera), files (\faFileO), and network transmission (\faWifi). 
``N/A'' shows that no native outputs are supported.
\textbf{*} Sepia supports cameras via extensions.
}
\label{tab:related}
\end{table}
Several manufacturers provide processing libraries that operate directly with their cameras, such as Prophesee's OpenEB~\cite{OpenEB_2022}, Inivation's libcaer~\cite{libcaer}, Celex's SDK~\cite{celex}, and jAER from the Institute of Neuroinformatics at the University of Zürich and the ETH Zürich~\cite{jaer}.
Recently, however, several independent libraries have been released for reading or processing AER data, including AEDAT~\cite{aedat_repo}, Expelliarmus~\cite{expelliarmus}, and Sepia (along with the processing library Tarsier)~\cite{Marcireau_Ieng_Benosman_2020}. 
Except from Sepia, they all offer Python interfaces, but are written in Rust, C, and C++ respectively.

Table \ref{tab:related} compares open-source libraries based on language, availability of Python bindings, and native support for inputs and outputs.
``N/A'' outputs shows that the library does not provide direct support for a peripheral output, but instead requires the user to program the integration themselves.
Note also that many of the file formats are different, and some have support for additional information such as inertial measurement units (IMUs) or even regular video.
But the formats are common in their support for some variation of AER data.

Since conventional signal processing algorithms cannot be applied to AER data, tailor-made algorithms have been developed for problems such as filtering, compression and feature extraction~\cite{Tayarani-Najaran_Schmuker_2021}.

\section{AEStream: Streaming address-event processing} \label{sec:aestream}
\begin{figure}
    \centering
    \includegraphics[width=0.7\textwidth]{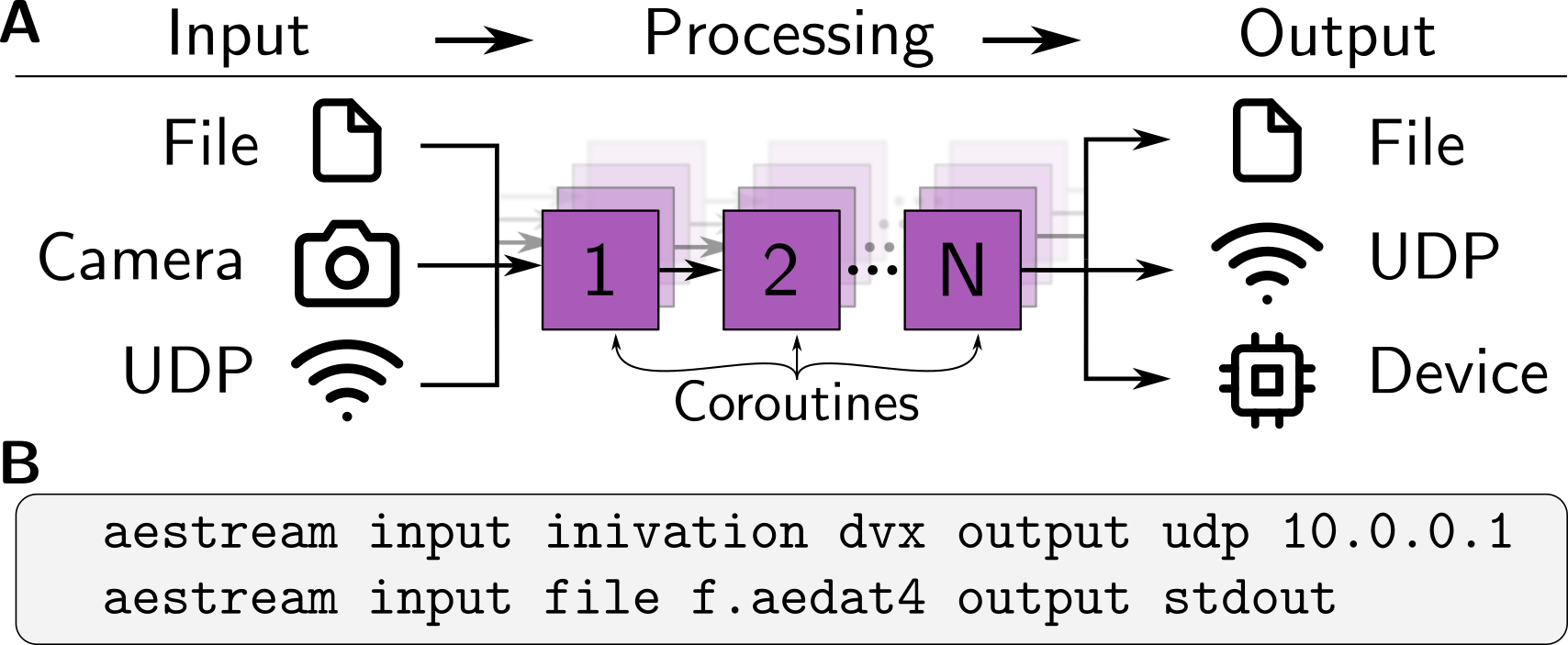}
    \caption{
    \textbf{(A)}~AEStream effectively streams address-event representations (AER) from input sources to output sinks via coroutines.
    \textbf{(B)}~Two examples from AEStream's command-line interface that illustrate the free composition of input-output pairs.
    }
  \label{fig:architecture}
\end{figure}

AEStream implements AER processing as coroutines in C++ operating on unbounded streams of events.
Instead of using on buffers or synchronized memory, AEStream operates directly on event-tuples $(x, y, p, t)$ which provides a strikingly simple architecture: functions of identical signatures can be freely combined to create the desired processing pipeline, as illustrated in Figure \ref{fig:architecture}.

In terms of inputs, AEStream supports EBV cameras from Inivation and Prophesee, as well as \texttt{.aedat4} files and network input via the SpiNNaker Peripheral Interface (SPIF) protocol \footnote{https://github.com/SpiNNakerManchester/spif} based on User Datagram Protocol (UDP) data.
In terms of outputs, AEStream supports \texttt{.aedat4} files, standard output, or network over UDP using the SPIF protocol, which has been used to stream data directly from events into the SpiNNaker neuromorphic hardware platform~\cite{Furber_Galluppi_Temple_Plana_2014}.

Apart from the command-line interface (CLI) shown in Figure \ref{fig:architecture} \textbf{(B)}, we also implemented Python wrappers such that events can be directly accessed as multidimensional arrays (tensors) in PyTorch~\cite{Paszke2019PyTorchAI}.
This allows seamless integration with PyTorch as well as tools within the machine learning ecosystem, such as the spiking neural network simulator, Norse~\cite{Pehle_Pedersen_2021}.
PyTorch and Norse heavily utilized GPUs to accelerate the computation, so we added native support for GPUs, to be explored further in Section \ref{sec:gpu} below.

In the remainder of the paper, we will first benchmark the impact of coroutines and then demonstrate an end-to-end application of AEStream, including code examples and performance numbers.

\subsection{Coroutine benchmarks}
\begin{figure}
    \centering
    \includegraphics[width=\textwidth]{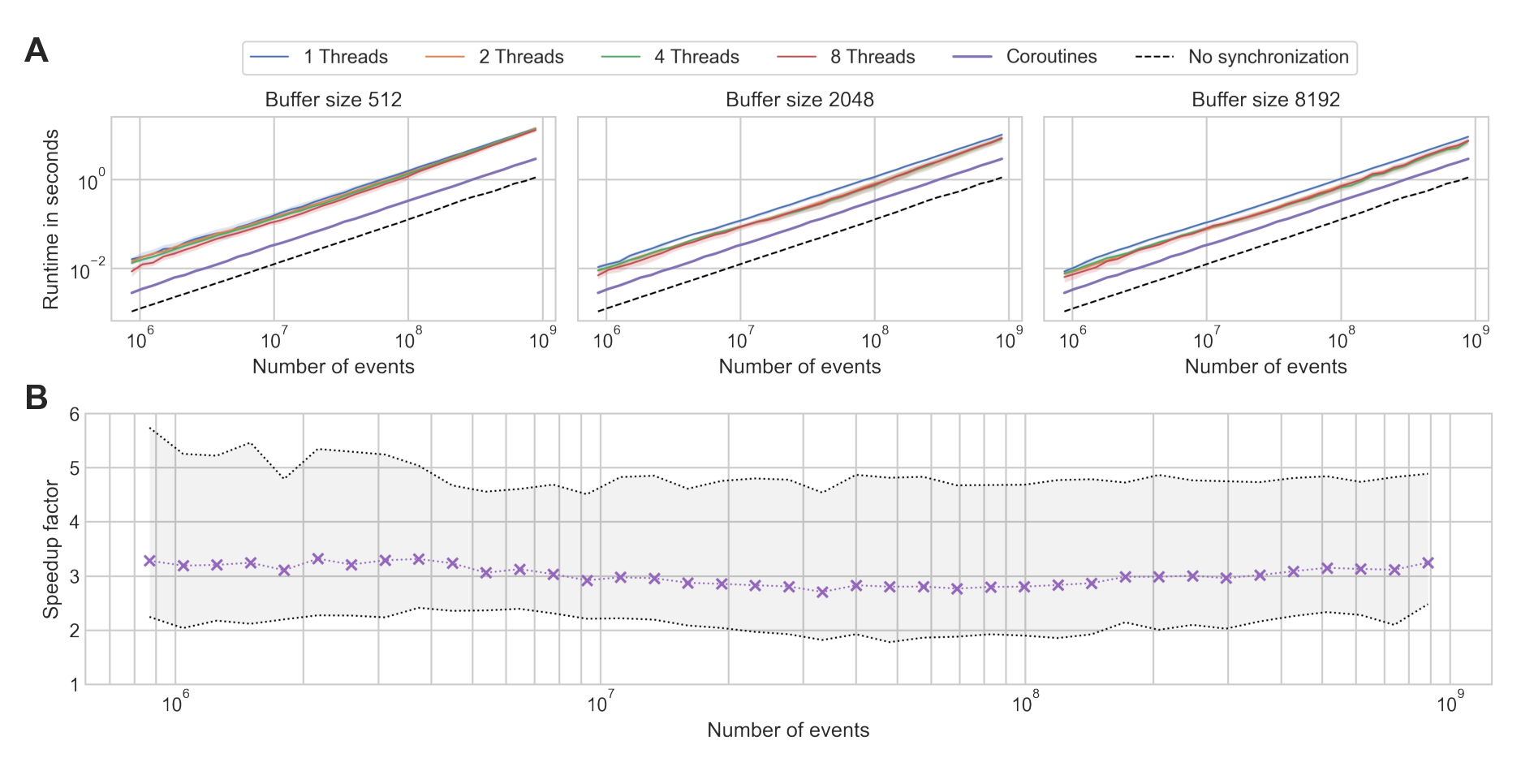}
    \caption{
    Relative throughput improvement of coroutines compared to threads for a trivial workload with a varying number of events. All numbers are averaged over 128 runs.
    \textbf{(A)} Runtime comparisons between threads, coroutines, and a simple function call without any threading or synchronization.
    \textbf{(B)} The purple line shows the \textit{relative} speedup of coroutines compared against the \textit{mean} runtime of threads.
    The black lines denote the interval between the \textit{minimum} and \textit{maximum} relative speedup.
    }
    \label{fig:speedup}
\end{figure}

To measure the actual impact of coroutines on processing performance, we devised a benchmark comparing coroutines to conventional thread programming.
Specifically, we wish to isolate the impact on asynchronous throughput, given the need for \textit{some} synchronization barrier as described in Section \ref{sec:sync}.
We, therefore, compared coroutines to a threaded approach, where one or more threads wait for fixed-size buffers to process.
To create the buffers, a single thread reads from a massive event array cached in random access memory (RAM) to avoid delays from disk I/O.
As a baseline comparison against the coroutines and threads, we also add a single-threaded non-synchronization method.

The actual work done in the benchmark is as straight-forward as possible to separate the effect of the synchronization: we simply sum up the coordinates in every event as a form of checksum that is verified against the true checksum at the end of the benchmark.
To avoid statistical effects, we repeat every step in the simulation 128 times.
Benchmarking was done on an AMD9 Ryzen Threadripper 2950X 16-Core CPU machine with 64GB RAM, running Ubuntu Linux with kernel 5.15.0-47.
Further details, along with the full source code for the benchmark, can be found online at \href{https://jegp.github.io/aestream-paper}{https://jegp.github.io/aestream-paper}.

Figure \ref{fig:speedup} part \textbf{(B)} shows the individual benchmarking runs with buffer sizes of $2^8$, $2^{10}$, and $2^{12}$.
Since coroutines do not use buffers, the purple line is identical across the charts.
The dashed black line indicates baseline performance without synchronization and is, therefore, also constant in all charts.
We observe that the runtime difference between coroutines and threads are relatively constant, despite the varying number of events being processed and the different buffer sizes.
\textbf{(A)} shows the average relative speedup between coroutines and threads.
The purple line indicates the speedup of coroutines compared to the \textit{averaged} runtime of threads across the various buffer sizes shown in \textbf{(B)}.
The purple shaded area shows 95\% of all the runs.
The two black lines indicate the speedup of coroutines compared to the \textit{minimum} and \textit{maximum} runtime of threads with varying buffer sizes.
Recall that the computational work in the benchmark is trivial, but purely in terms of concurrency the results are promising: coroutines provide at last 2 times higher throughput compared to conventional threads, irrespective of buffer sizes and number of threads.

\section{Use case: edge-detection with GPUs} \label{sec:gpu}
\begin{wrapfigure}{R}{0.55\textwidth}
    \centering
    \includegraphics[trim=20 10 30 10,width=0.5\textwidth]{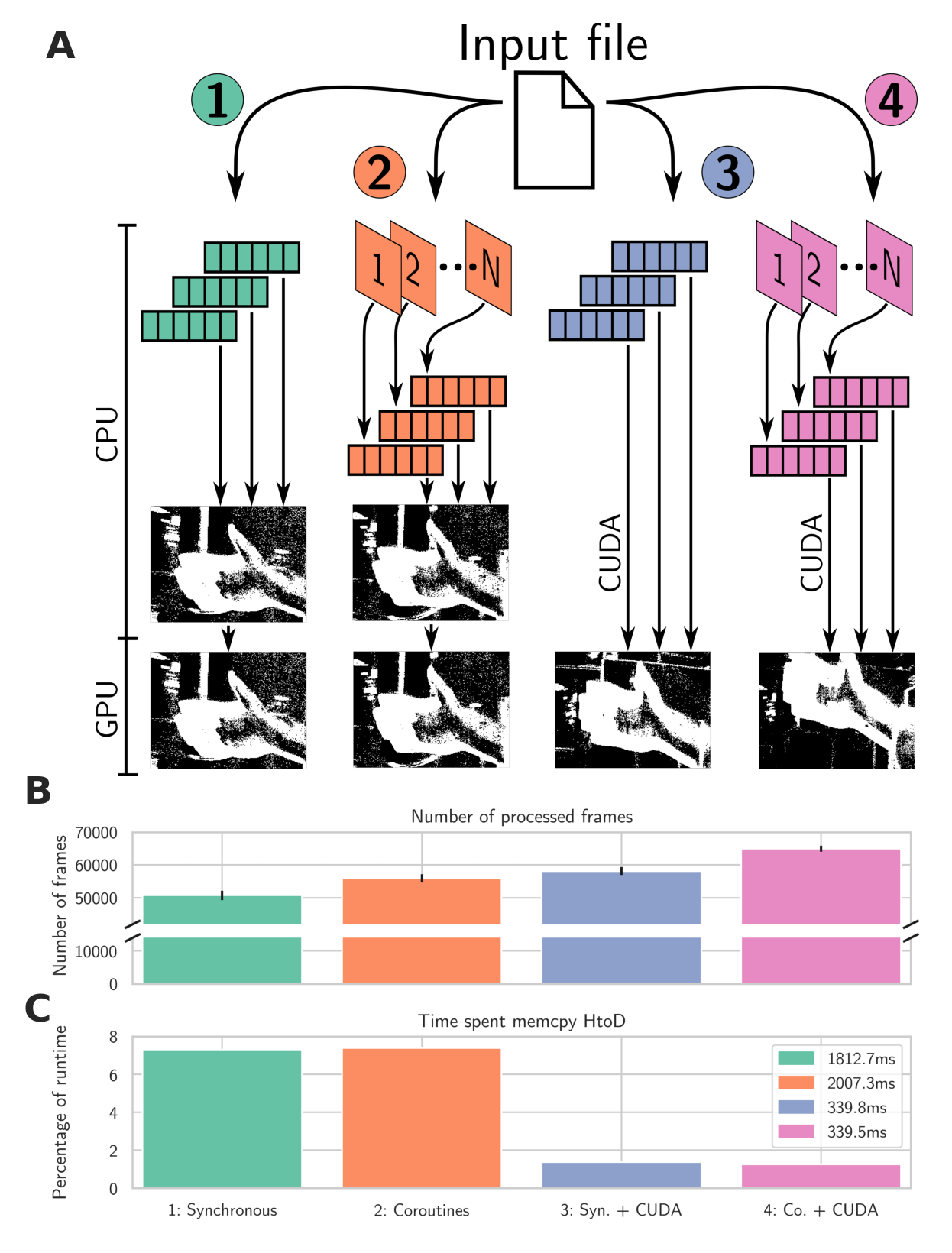}
    \caption{
    \textbf{(A)} In our experiment, events are carried from a file via the CPU to the GPU in four different ways.
    \textbf{(B)} Time spent copying memory from host to device (HtoD) shown as a percentage of the total runtime as well as in milliseconds.
    \textbf{(C)} The number of frames that were run through the edge detector during the benchmark.
}
    \label{fig:gpu_threads}
\end{wrapfigure}
We proceed to demonstrate AEStream in a practical setting: detecting edge-detection from AER data in real-time with a GPU.
Our aim is twofold: (1) to document the benefits of parallel I/O as sketched above, and (2) to demonstrate how AEStream can connect directly with specialized compute peripherals, such as GPUs.


The specific setting is shown in Figure \ref{fig:gpu_threads} \textbf{(A)} where an input file is streamed to a GPU that uses a spiking neural network to detect edges in the incoming events.
Spiking neural networks are biologically inspired neural networks that operate with events, similar to the AER schema~\cite{Maass_1997}.
Our network combines a leaky integrate-and-fire (LIF) neuron layer (with an added refractory term to reduce noise) and a regular convolution.
We use the Norse simulator~\cite{Pehle_Pedersen_2021} to build and evaluate the model, which allows us to port it directly to the GPU.

Norse operates on tensors, which requires us to bin our events into ``frames''.
This can be achieved in two ways: we can either generate the tensor on the CPU and then copy it over, or we can copy the AER data over and generate the tensor on the GPU.
The latter approach grants two advantages: (1) it reduces the amount of data we transfer to the GPU because we can leverage the sparsity of AER data compared to the (mostly empty) large matrix representation.
In turn, (2) we reduce the CPU load because we can defer processing to the GPU.
This is particularly desirable for GPUs because they permit simultaneous operation on thousands of events using the Compute Unified Device Architecture (CUDA) by NVIDIA~\cite{Nickolls2008ScalablePP}.

Parameterizing our setup over (1) coroutines and (2) the use of CUDA kernels to copy over data, we arrive at four different scenarios, as shown in Figure \ref{fig:gpu_threads} \textbf{(A)}.
In the first scenario, a single thread synchronously populates several buffers from a file. 
When a buffer is full, the thread copies the buffers directly onto a CPU tensor that is copied to the GPU and input into the edge detector.
The second scenario uses coroutines when processing the events and filling the buffers, but still copies the entire tensor to the GPU.
In the third scenario, we exploit the above-mentioned CUDA kernels to parallelize the transfer of AER data to the GPU that then applies it to the edge detector.
In the fourth scenario, we use coroutines to populate buffers that are then transferred to the GPU using the same CUDA kernels as in the third scenario.

\subsection{Experimental setup}
\begin{figure}
    \centering
\begin{lstlisting}[language=Python]
import aestream
with aestream.FileInput("file.aedat4", device="gpu") as file:
    while True:         # Loop as fast as possible
        t = file.read() # Grab a tensor for further processing
\end{lstlisting}
    \caption{
    Python code to stream a file to a GPU, such that they later can be \texttt{read()} in the form of a PyTorch tensor.
    }
    \label{fig:file_code}
\end{figure}

Practically speaking, we are streaming a file with 90 million events recorded for 24.8 seconds real-time from a $346\times260$ resolution camera (select frames from the recording are shown in Figure \ref{fig:gpu_threads} \textbf{(A)}).
Figure \ref{fig:file_code} shows the code for opening an event-based file and streaming it to a GPU.
When filling the buffers, we respect the timestamps in the file, meaning that all our benchmarks will last at least 24.8 seconds.
However, we are \textit{not} limiting the number of tensors the GPU can process per second.
Since the computation in all programs are constant, this allows us to focus on what happens \textit{after} the synchronization barrier towards the GPU.
Specifically, we measure (1) how many tensor ``frames'' the GPU can process per second in the different settings and (2) how much time we are spending copying memory from the host (CPU) to the device (GPU).
The full version of the benchmark code, along with everything required to reproduce our results, is available in the project repository.

\subsection{Results}
Figure \ref{fig:gpu_threads} \textbf{(B)} reveals that it is significantly faster to copy AER data to the GPU compared to copying full tensors.
In terms of the entire $> 24.8$ seconds of the benchmark, the first two scenarios spend around 7\% of the entire runtime copying data to the GPU, while the other two scenarios use slightly less than 2\%.
Since coroutines are not involved with copying memory, it is expected that they have little effect on performance.

Part \textbf{(C)} shows the number of processed frames over the course of the benchmark.
Recall that there is no bound on the GPU part of the benchmark, so the programs are free to grab as many frames as possible per second.
In the figure, there are clear effects of both the coroutines and the custom CUDA kernels.
As we show above, the concurrency inherent in coroutines reduces the amount of time spent on ``hot spots'', which in this case means locks around the buffers feeding events into the tensors.
That is a write operation into the tensors, so the lock prevents the user from taking a new tensor and giving it to the spiking neuron model.
CUDA kernels also have a clear effect, which is related to the reduction in the copying operations in plot \textbf{(B)} because the GPU spends less time waiting for memory.

In sum, AEStream reduces copying operations to the GPU by a factor of at least 5 and allows processing at around 30\% more frames ($6.5\times 10^4$ versus $5\times10^4$ in total) over a period of around 25 seconds compared to conventional synchronous processing.

\section{Discussion}
We discussed the efficient processing of events on von Neumann computers using coroutines and implemented efficient concurrent processing of address-event representations (AER) in the library AEStream.
AEStream is designed to read, process, and emit events at high speeds to avoid bottleneck effects when operating high-bandwidth neuromorphic sensors.
Because AEStream is built around simple, functional primitives, it can arbitrarily connect inputs to outputs in a straight-forward Python or command-line interface.
This is particularly useful for real-time settings where dedicated, parallel hardware, is needed to process the deluge of incoming events.
Thanks to the SpiNNaker Peripheral Interface (SPIF) protocol, connecting an event-based camera with SpiNNaker can be done with one command in AEStream.

Another important contribution is the PyTorch and CUDA support.
Low-level CUDA programming is a non-trivial and error-prone endeavor that is rarely worth the effort.
In the case of GPU-integration, however, we believe the effort is timely; efficiently sending AER data to GPUs opens the door to a host of contemporary machine learning tools, ranging from deep learning libraries like PyTorch \cite{Paszke2019PyTorchAI} to spiking neuron simulators like Norse \cite{Pehle_Pedersen_2021} to graphical libraries for visual inspection.

\paragraph{Limitations}
It should be said that we have not studied AEStream in relation to other libraries, so we can only hypothesize how fast or slow it performs in comparison.
Further benchmarks in this direction would be interesting.

When sending AER data to GPUs we are presently relying on dense tensor representations. 
Sparse tensors have recently been introduced in the PyTorch and machine learning ecosystems, and, although they are still not widely adopted, sparse tensors could greatly benefit the neuromorphic community because they remove significant overhead that is unnecessary in AER regimes.

\paragraph{Future work}
We are, ourselves, eager to apply AEStream to process events in real-time with neuromorphic hardware.
Eventually, we aim to stream events \textit{back} to an actuator to create a closed-loop, fully neuromorphic control system in real-time.

Due to the many possible permutations and combinations of inputs and outputs, AEStream is also well suited for multimodal sensing and sensor fusion.
Sending multiple inputs to a single neuromorphic compute platform would, for instance, be trivial.

More work is needed to explore further concurrency and parallelism benefits with AEStream. 
Particularly system-specific benchmarks would be important.
We have had success deploying AEStream on embedded systems, but there is presently no guarantee that bottlenecks do not occur.

It is our hope that AEStream can benefit the community and lower the entrance-barrier for research in neuromorphic computation.
Our code, along with instructions on how to reproduce our results, are openly available at \href{https://jegp.github.io/aestream-paper}{https://jegp.github.io/aestream-paper}.

\section{Acknowledgments}
We foremost would like to thank Anders Bo Sørensen for his friendly and invaluable help with CUDA and GPU profiling.
Without the thoughtful and extensive comments by Gregor Lenz, this paper would be much harder to read. 
Emil Jansson deserves our gratitude for scrutinizing and improving the coroutine benchmark C++ code.
We gracefully recognize funding from the EC Horizon 2020 Framework Programme under Grant Agreements 785907 and 945539 (HBP).
Our thanks also extend to the Pioneer Centre for AI, under the Danish National Research Foundation grant number P1, for hosting us.

\bibliography{references}

\end{document}